\documentclass[%
 reprint,
 amsmath,amssymb,
 aps,
]{revtex4-1}

\usepackage{graphicx}
\usepackage{dcolumn}
\usepackage{bm}
\usepackage[colorlinks=true,linkcolor=blue,urlcolor=blue,citecolor=blue]{hyperref}
\usepackage{url}

\begin{document}
\preprint{APS/123-QED}

\author{Chong Wang}
\author{Xiaofeng Ren}
\author{Yanxiang Zhao}
\affiliation{Department of Mathematics, George Washington University \\
Washington, DC 20052}

\title{Bubble Assemblies in Ternary Systems with Long Range Interaction}

\begin{abstract}
  A nonlocal diffuse interface model is used to study bubble assemblies
  in ternary systems.  As model parameters vary, a large number of morphological phases appear
as stable stationary states. 
One open question related to the polarity direction of double bubble assemblies is answered numerically.
Moreover, the average size of bubbles in a single bubble assembly
depends on  the sum of the minority constituent areas and the long
range interaction coefficients. One identifies the ranges for area
fractions and the long range interaction 
coefficients for double bubble assemblies.

\end{abstract}

\maketitle

{\it{Introduction.}}---Block copolymers have generated much interest in materials science in recent years due to their remarkable ability for self-assembly into nanoscale ordered structures \cite{block, developments, optoelectronics}.
This ability can be exploited to create materials with desired mechanical, optical, electrical, and magnetic properties \cite{block, developments, optoelectronics}. 
There have been many experimental and theoretical studies focusing on this subject \cite{  takenaka, blockCoTheo4, stable, drolet,  catyler, discoveringOrd, nucleationOrd,  jiang, Oono, cell}.
Self-consistent field theory derived from a microscopic description of interacting polymer chains is one successful theoretical approach for the study of block copolymers
\cite{ blockCoTheo4, stable, drolet, catyler, discoveringOrd, nucleationOrd, jiang}.  
However, this method is computationally demanding because of the heavy calculation of path integrals for the chain conformation \cite{cellDy, phase-field}.
There is a need for efficient methods to model the self-assembly for block copolymers at the mesoscale level. The density functional theory (DFT) \cite{equilibrium, microphase} is a very promising approach to modeling such phenomena and it is customarily referred to as cell dynamics simulation \cite{Oono, cell}.

In this Letter, we consider a computational model which originates from DFT  for block copolymers \cite{microphase} and has been extensively used to study morphological phases of diblock copolymers both analytically \cite{equilibrium, kineticPath, some, ontheMulti, onthe, variation, dropletPhases, phase-field, reorientation} and numerically \cite{cell, aPreconditioner, amixed, efficient}. 
Considerably less theoretical work exists for triblock copolymers. The symmetric triblock system
was studied using DFT in 
\cite{microphase, bohbot}. The three-component triangle phase diagram for triblock copolymers was reported based on DFT in \cite{morphologyABC}. The existence of core-shell assemblies was theoretically established in \cite{coreshell}.
The ternary systems studied here include triblock copolymers and homopolymer/diblock copolymer blends, in which two constituents have smaller volume fractions (area fractions in two dimensions)
compared to the third constituent. This study focuses on two dimensional morphological patterns where the minority constituents form assemblies of discs, called single bubbles, or assemblies of double bubbles, or assemblies of coexisting single and double bubbles.

In double bubble assemblies,
a two-thirds power law between the number of double bubbles and the long range interaction coefficients in the strong segregation regime is justified both numerically and theoretically.
A range of parameters is identified that yields double bubble assemblies.
In single bubble assemblies, it is shown that the average size of bubbles does no depend on the ratio of the area fractions but rather is determined by the sum of the minority constituent areas and the long range interaction coefficients.

{\it{Model summary.}}---Consider a free energy model for $ABC$ ternary systems originally derived by Nakazawa and Ohta for triblock copolymers  \cite{microphase}: 
\begin{align} \label{energyE}
E (\phi_1, \phi_2)  
&=   \int_{\Omega} \Big[  \frac{\epsilon}{2}  \big ( |\nabla \phi_1|^2   +   |\nabla \phi_2|^2   +  \nabla \phi_1 \cdot \nabla \phi_2  \big)  \nonumber \\
&   \qquad \qquad +   \frac{1}{2\epsilon}  W_{T}(\phi_1, \phi_2 ) \Big ] dx \nonumber\\
&+    \sum_{i,j = 1}^2 \frac{\gamma_{ij}}{2}  \int_{\Omega} \Big [ \left(-\triangle\right)^{-\frac{1}{2}}  \left( f(\phi_{i}) -\omega_{i}\right) \times \nonumber \\
 & \qquad \qquad \qquad  \left(-\triangle\right)^{-\frac{1}{2}}\left( f(\phi_{j}) -\omega_{j}\right) \Big] dx.  
\end{align}
Here $\Omega$ is a bounded domain in $\mathbb{R}^d$, $d=1$,  $2$ or $3$. $\phi_i, i = 1, 2$ are phase-field labeling functions which represent relative constituent (e.g., monomer) density fields.
$ \{x: \phi_i(x) \approx 1 \}$, $i = 1, 2$ stand for the regions with high concentration in $A$ and $B$ constituents, respectively.
  The concentration of $C$-constituent can be implicitly represented by $1 - \phi_1(x) - \phi_2(x) $ since the system is assumed to be incompressible \cite{morphologyABC}.
Here the interfacial thickness $\epsilon \ll 1$; the system is in the strong segregation regime \cite{cell}. 
$W_T(\phi_1,\phi_2) $ is of the form of
\begin{eqnarray}
W_T(\phi_1,\phi_2) :=  W(\phi_1)  + W(\phi_2) + W(1 -  \phi_1 - \phi_2), \nonumber 
\end{eqnarray}
where $
  W(\phi) = 18 (\phi^2-\phi)^2. 
  $
Note that $W_T(\phi_1,\phi_2)$ is a triple-well potential which has three minima at $(1,0,0)$, $(0,1,0)$ and $(0,0,1)$.
The first integral in \eqref{energyE} describes the short range interaction which accounts for the interfacial free energy of the system and favors large domains with minimum surface area.

The long range interaction coefficients $\gamma_{ij}$ form a symmetric two
by two matrix $\gamma = [\gamma_{ij}]$. 
For triblock copolymers, the matrix $\gamma$ is positive definite \cite{triblockCo}; for homopolymer/diblock copolymer blends $\gamma$ has one positive
eigenvalue and one zero eigenvalue \cite{blend}.
 This work studies the effect of $\gamma$ in a wide range, including positive definite and non-positive definite cases.
Define $ \mathring{L}^2 = \{ g \in L^2(\Omega) : \int_{\Omega} g dx = 0 \}$, square integrable functions with zero mean, and $ \mathring{H}^2 = \{ v \in H^2(\Omega) : \int_{\Omega} v dx = 0 \}$.
  The inverse negative Laplace operator 
  $(- \triangle )^{-1}$:  $\mathring{L}^2 \mapsto \mathring{H}^2 $ is defined by
  \begin{eqnarray} 
 ( - \triangle )^{-1} g = v  \; \text{ iff}   - \triangle v = g, \quad  g \in \mathring{L}^2, \; v \in \mathring{H}^2, \nonumber
 \end{eqnarray}
  with periodic or homogeneous Neumann boundary conditions.
  The nonlocal operator $(- \triangle)^{-\frac{1}{2}}$ is its positive square root.
The function $f(\phi_i)$ satisfies the following conditions:
\begin{eqnarray}
f(\phi_i = 0) = 0, \; f(\phi_i = 1) = 1, \nonumber \\
f^{\prime}(\phi_i = 0) = 0, \; f^{\prime}(\phi_i = 1) = 0. \nonumber
\end{eqnarray}
Introducing such a function $f$ will localize the force near the interface and avoid the possible unphysical feature of negative values in relative constituent (e.g., monomer) density fields \cite{newPhasef}.
 Here we choose 
\begin{eqnarray} \label{fphi}
f(\phi_i) = (\phi_i^2 - 2\phi_i)^2, \quad  i=1,2.
\end{eqnarray}
In \eqref{energyE} $\omega_i$ and $\gamma_{ij}$ are the main parameters.
One imposes the constraints
\begin{equation}
  \omega_i = \frac{1}{|\Omega|} \int_{\Omega} f( \phi_i  )  dx,
  \label{constr}
\end{equation}
so $\omega_i$ denotes the volume fraction of  the $i$-th constituent.
The $\gamma_{ij}$ terms in \eqref{energyE} describe the long range interaction which accounts for the chain conformation energy \cite{morphologyABC} and favors small domains.

To minimize the free energy \eqref{energyE}, consider the $L^2$  gradient flow dynamics 
\begin{eqnarray} \label{nonlocalAC}
\frac{\partial \phi_i}{\partial t} & = &  \epsilon \triangle \phi_i + \frac{\epsilon}{2} \triangle \phi_j  - \frac{1}{2\epsilon} \frac{\partial W_T}{\partial \phi_i}\nonumber \\
&& - \gamma_{ii} (- \triangle)^{-1} (f(\phi_i) - \omega_i)  f^{\prime}(\phi_i) \nonumber \\
&& - \gamma_{ij} (- \triangle)^{-1} ( f(\phi_j) - \omega_j)  f^{\prime}(\phi_i) ,  
\end{eqnarray}
in which $i, j = 1, 2 $ and $j \neq i$, coupled with the constraints
\eqref{constr}.

{\it{Numerical method.}}---Take $\Omega$ to be a square with the
periodic boundary conditions.
To meet the constraints \eqref{constr}, we adopt the modified augmented Lagrange multiplier approach \cite{multiplier,efficientSta}. This method on the one hand makes the system less stiff and on the other hand leads to the well posedness of $(- \triangle)^{-1} (f(\phi_i) - \omega_i)$.
Eqs. \eqref{nonlocalAC} coupled with \eqref{constr} can be solved efficiently by semi-implicit Fourier spectral method on a uniformly discretized spatial domain  \cite{applications, coarsening}.
Numerical simulations start from random initial configurations satisfying the constraints \eqref{constr}.

\begin{figure}
\includegraphics[width=3.8cm]{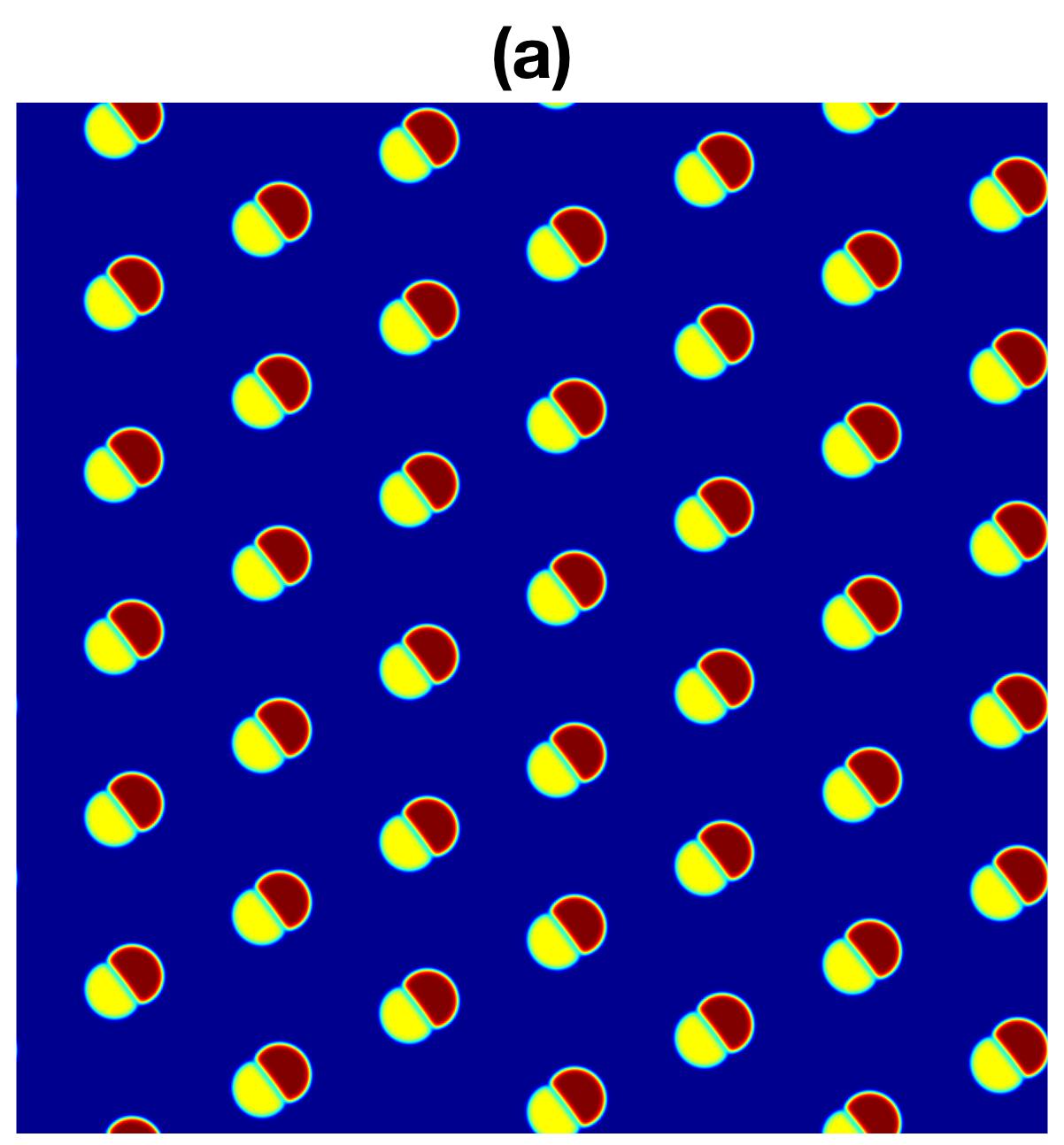} \quad
\includegraphics[width=3.8cm]{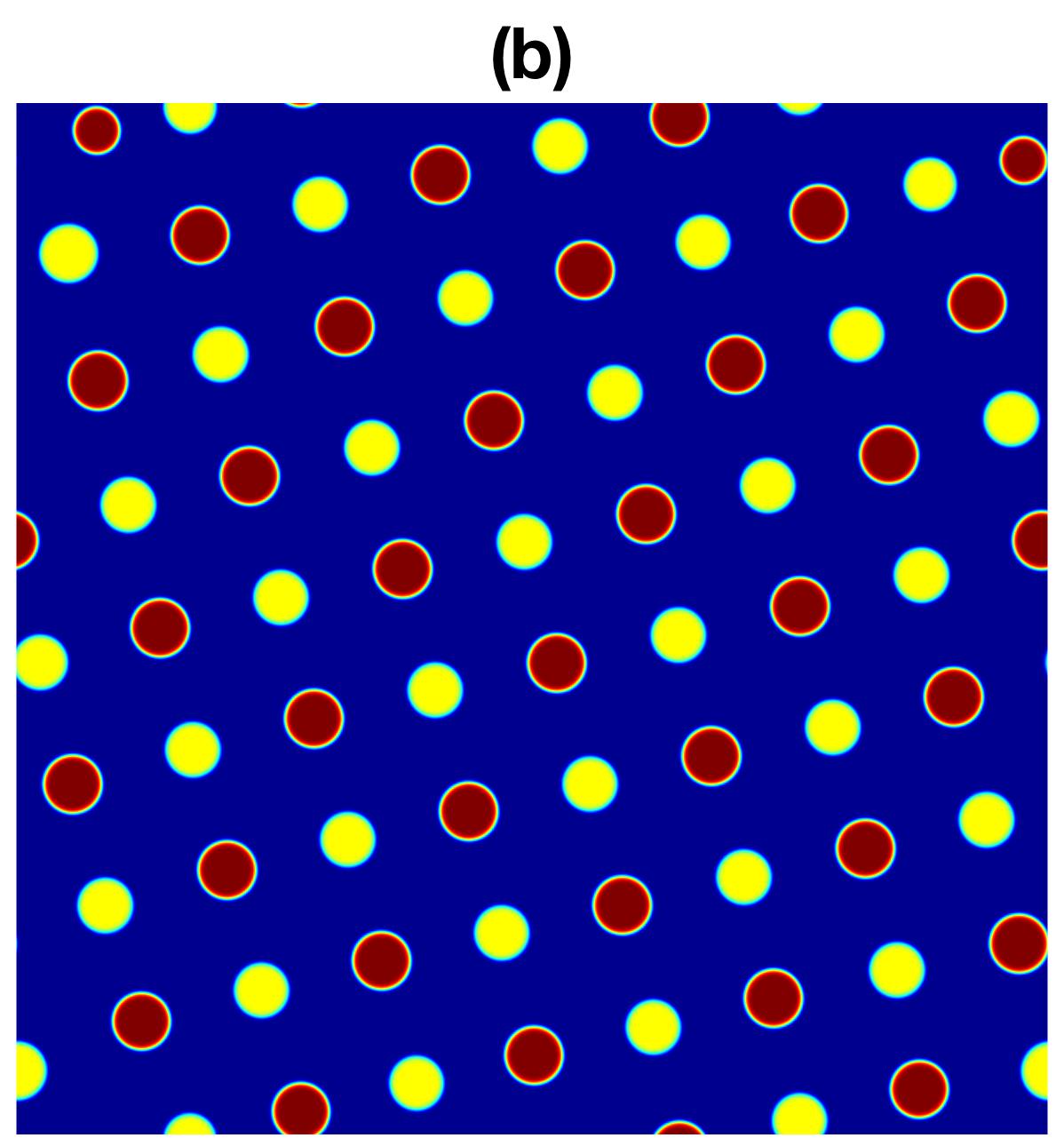} 
\caption{Two characteristic patterns in ternary systems. (a) A ternary system with $\gamma_{12} = 0$ maintains a hexagonal double bubble assembly. (b) A system with $\gamma_{12} = 11,000 $ yields a single bubble assembly in a square lattice. The snapshots are taken at time $T=400$.
  $\gamma_{11} = \gamma_{22} = 20,000$, $\omega_1 =0.10$, and $\omega_{2} = 0.09$ in these simulations.    
}\label{fig1}
\end{figure}

The five parameters, $\gamma_{11}$, $\gamma_{12}$, $\gamma_{22}$, $\omega_1$, and $\omega_2$, play the key roles in pattern formation of ternary systems. 
In numerical simulations, the domain $\Omega$ is fixed as $ [-1, 1]^2$, the uniform mesh grid in space is fixed as $512 \times 512$, namely, $\Delta x = \Delta y = 2/ 512 $,  $\epsilon $ is fixed as $5  \Delta x$, and the time step $\Delta t $ is  $ 0. 001 $. In each image below, red, yellow and blue colors correspond to 
$A$-rich, $B$-rich and $C$-rich regions, respectively.

\begin{figure}
\includegraphics[width=9cm]{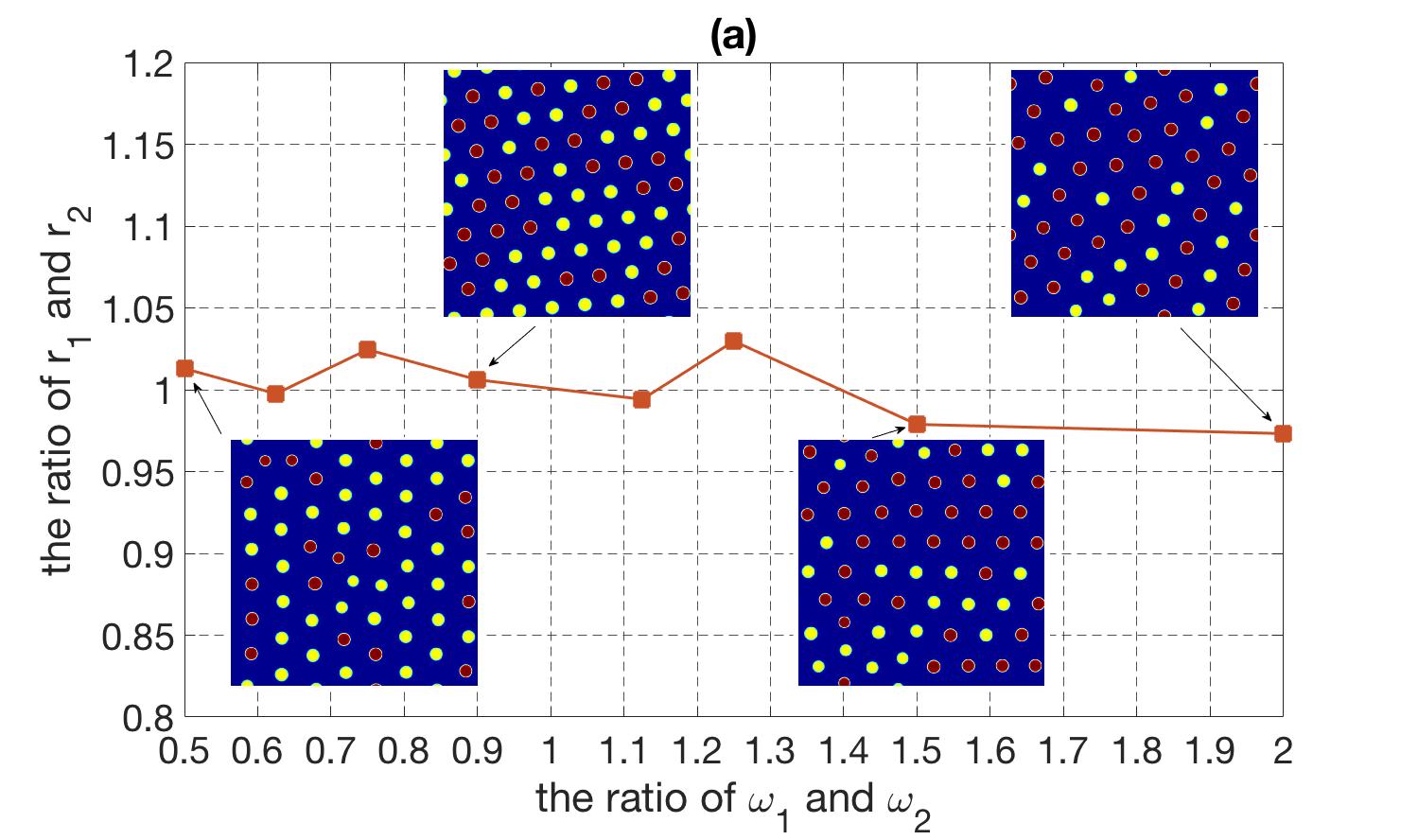}\\
\includegraphics[width=9cm]{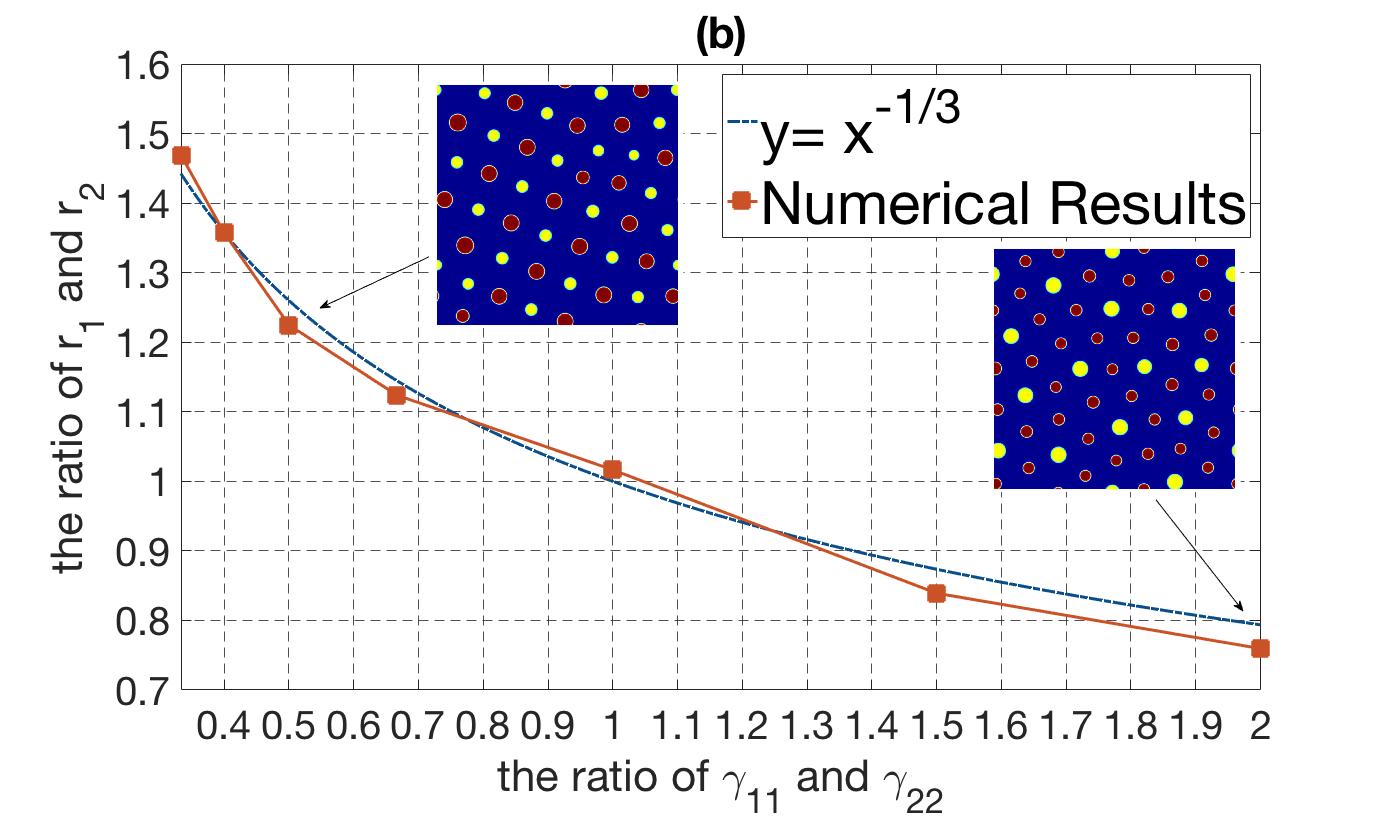}
\caption{(a) Independence of the average red and yellow bubble sizes on the ratio of area fractions $\omega_1/\omega_2$. For $(\omega_1, \omega_2) = (0.05, 0.10)$, $(0.09, 0.144)$, $(0.09,0.12)$, $(0.09, 0.10)$, $(0.09, 0.08)$, $(0.09, 0.072)$, $(0.09,0.06)$, $(0.10, 0.05)$, the ratio $r_1 / r_2$ remains at $1/1$ up to a $3  \% $ error. Here $\gamma_{11} =
\gamma_{12} = \gamma_{22} = 20, 000$.
(b) Dependence of the average red and yellow bubble sizes on the long range interaction coefficients $\gamma_{11}$ and $\gamma_{22}$. For $(\gamma_{11}, \gamma_{22} ) = (20,000, \ 60,000)$,  $(20,000, \ 50,000)$, $(10,000, \ 20,000)$, $(20,000, \ 30,000)$, $(20,000, \ 20,000)$, $(30,000, \ 20,000)$, $(20,000, \ 10,000)$, numerical simulations agree with the law of $r_1/r_2 = (\gamma_{11}/\gamma_{22})^{-1/3}$.
Here $\gamma_{12} = 20, 000, \ 20, 000, \ 10, 000, \ 20, 000, \ 10, 000, \ 20, 000, \ 10, 000$ respectively.
$(\omega_1, \omega_2) = (0.10, 0.05)$, $(0,10, 0.05)$, $(0.09, 0.06)$, $(0.10, 0.05)$, $(0.07, 0.07)$, $(0.09, 0.06)$, $(0.09, 0.06)$ respectively.
}
\label{fig2}
\end{figure}

{\it{Sample equilibria.}}---Two sample equilibria are presented in Fig. \ref{fig1}. 
Fig. \ref{fig1} (a) shows a double bubble assembly. 
All double bubbles grow into the same size and are located hexagonally. 
The polarity direction of each double bubble, the direction from center of mass of yellow region to that of red one, in an assembly is unknown theoretically
\cite{double}.  Numerical simulations show double bubble assemblies when $|\gamma_{12}|$ is small, and the polarity directions of double bubbles in equilibrium configurations are parallel.
Fig. \ref{fig1} (b) shows a single bubble assembly.
All yellow bubbles become equal in size, as do red bubbles. Interestingly, they form a square lattice pattern in which each single bubble is surrounded by four bubbles of the other color. In a binary system, a hexagon pattern is most stable experimentally \cite{molecularCon} and theoretically \cite{blockCoTheo6, theory, equilibrium, appliModular, many}. For a ternary system, our numerical simulations show that a square structure can be energetically more favorable than a hexagonal one.
This agrees with experiments \cite{squarePacking} and theoretical studies \cite{microphase, morphologyABC, pingtang}.

\begin{figure}
\includegraphics[width=9cm]{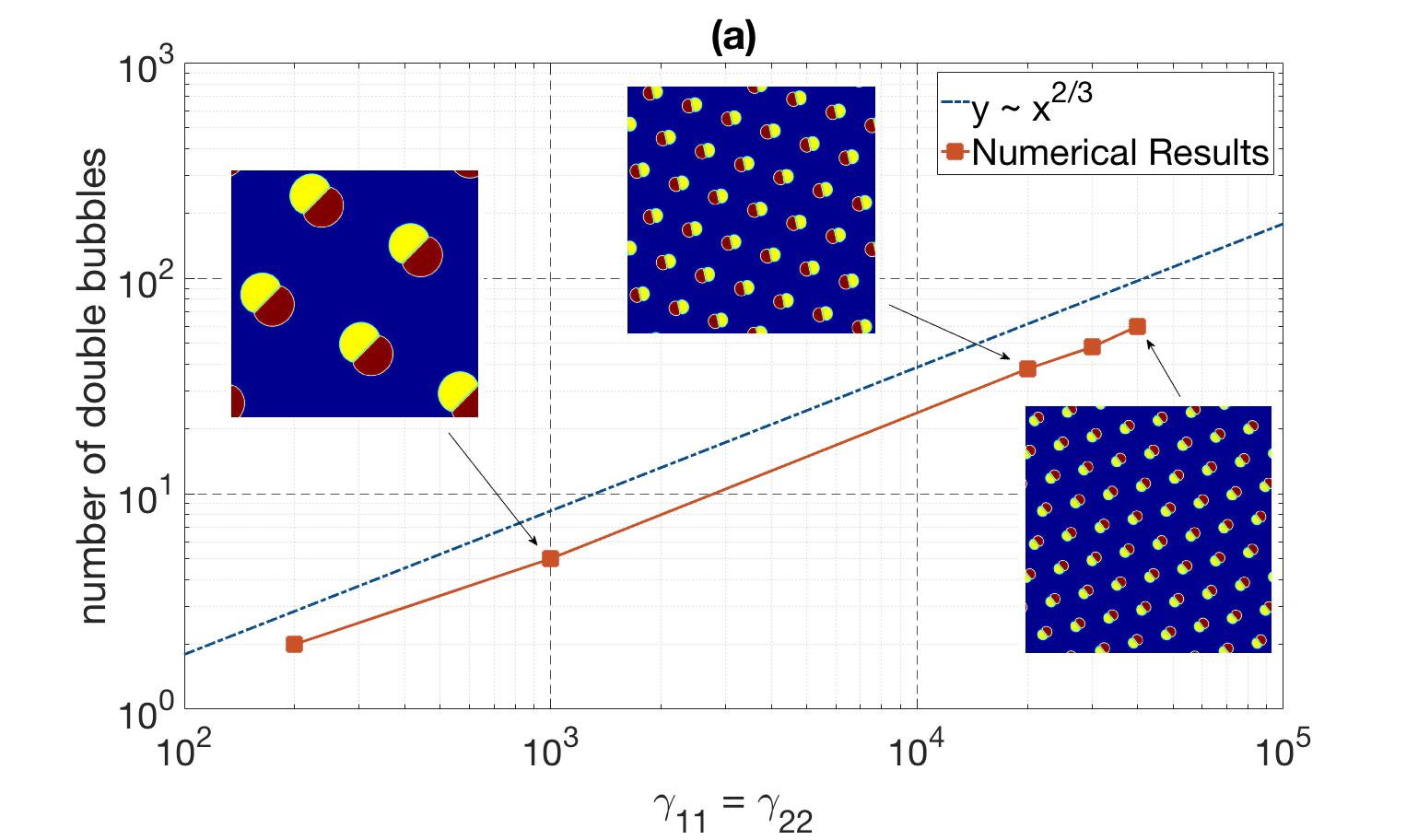} \\
\includegraphics[width=9cm]{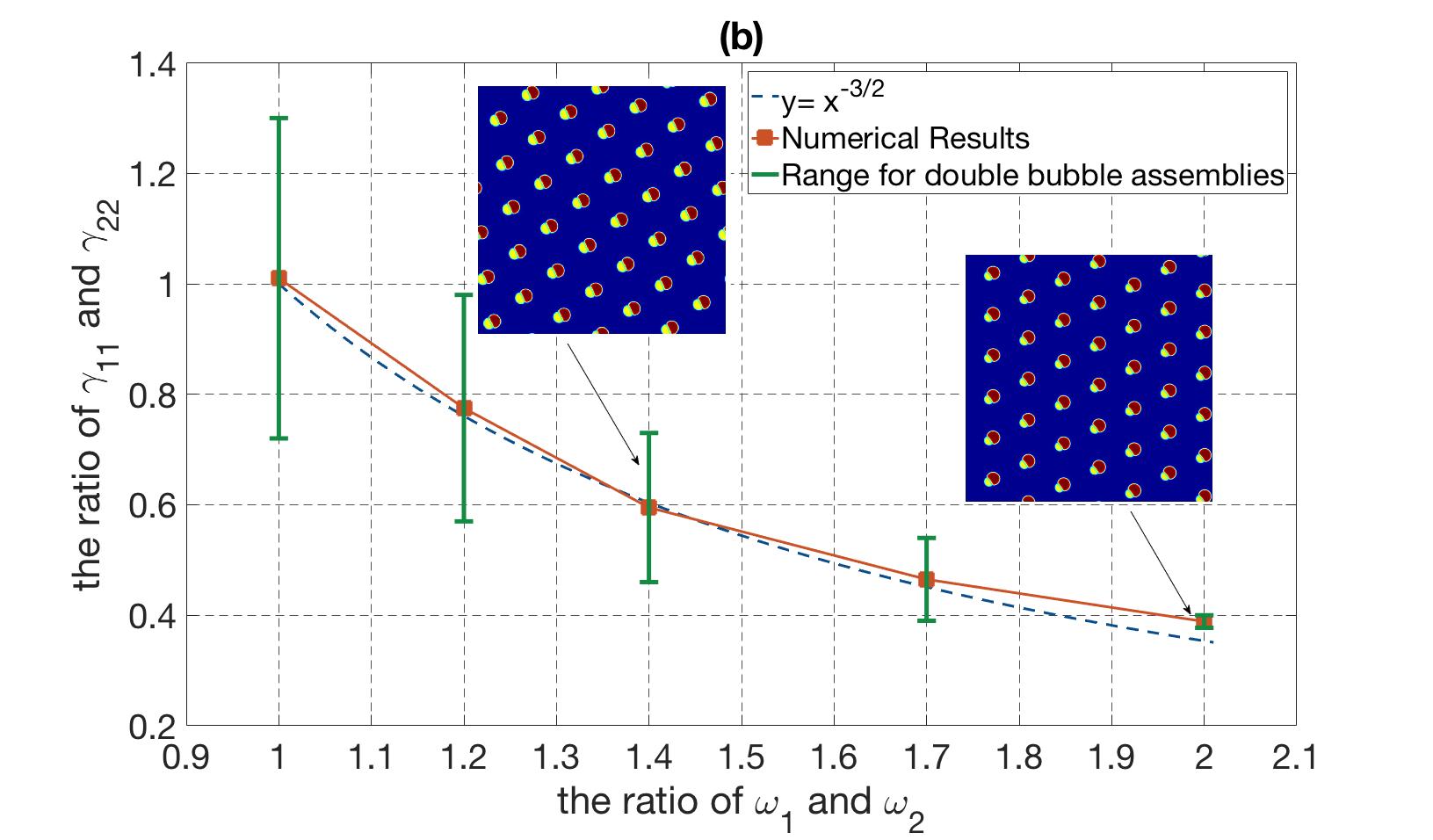}
\caption{ (a) Log-log plot of the dependence of the number of double bubbles on $\gamma_{11}$ in symmetric double bubble assemblies. Here $\gamma_{11} = \gamma_{22}$, $\gamma_{12}=0$, and $\omega_1 = \omega_2=0.09$. As $\gamma_{11}$ increases, the number of double bubbles in the assemblies grows accordingly.
For $\gamma_{11}= 200$, $1,000$, $20,000$, $30,000$, $40,000$, the corresponding number of double bubbles are $2$, $5$, $38$, $48$ and $60$, respectively. 
(b) The range of $ \gamma_{11}/ \gamma_{22} $ under which random initials evolve to double bubble assemblies for given $\omega_{1} $ and $ \omega_{2} $.  
For $(\omega_1, \omega_2) = (0.09, 0.09)$, $(0.09, 0.09/1.2)$, $(0.09, 0.09/1.4)$, $(0.10, 0.10/1.7)$, $(0.10, 0.05)$, the ranges of $\gamma_{11} / \gamma_{22}$ are $(0.72, 1.3)$, $(0.57, 0.98)$, $(0.46, 0.73)$,  $(0.39, 0.54)$, and $(0.3774, 0.40)$ respectively. Here $\gamma_{12} = 0. $} 
\label{fig3}
\end{figure}

{\it{Single bubble assemblies.}}---For single bubble assemblies,  the average size of red/yellow bubbles does not depend on the ratio of area fractions, namely, $\omega_1/\omega_2$.
In Fig. \ref{fig2} (a), for several $(\omega_1, \omega_2)$ and fixed $\gamma_{ij} = 20,000, \ 1 \leq i,j \leq 2$, the ratio $r_1/r_2$ remains at 1/1 up to a 3\% error regardless of the different values of $\omega_1/ \omega_2 $. Note that $(\omega_1, \omega_2)$ has an impact on the number of red/yellow bubbles, as seen in the insets of  Fig. \ref{fig2} (a).
On the other hand, the values of $\gamma_{11}$ and $\gamma_{22}$ affect $r_1/r_2$.
In Fig. \ref{fig2} (b), with various $(\gamma_{11},\gamma_{22})$, the ratio $r_1/r_2$ decreases as $\gamma_{11}/\gamma_{22}$ becomes larger. More precisely, the two ratios satisfy the following law:
 \begin{align} \label{formulaR1R2}
\frac{r_1}{r_2}= \left ( \frac{\gamma_{11}}{\gamma_{22}} \right)^{-\frac{1}{3}} .
 \end{align}
 
 This relationship can also be verified theoretically.
 Consider the strong segregation
 limit (known as the $\Gamma$-limit in mathematics)
 of the free energy $E$ \cite{minimal}.
Let $K_1$ be the number of red bubbles and $K_2$ be the number of yellow bubbles in a single bubble assembly.
In an equilibrium state, all red bubbles develop into approximately the same size; so do yellow bubbles. 
Let $r_1$ and $r_2$ be the average radii of red and yellow bubbles, respectively. Based on \cite{discAssemblies}, up to the leading order, the free energy is 
\begin{eqnarray} \label{newEsingle1}
\sum_{i=1}^2 K_i \left (2 \pi r_i  +  \frac{\gamma_{ii} \pi}{4} (r_i)^4 \log \frac{1}{r_i} \right ). 
\end{eqnarray}
Let $\eta^2 m = \omega_1 |\Omega|$,  $\eta^2 (1 - m ) =  \omega_2 |\Omega|$, and $\Gamma_{ij} = \eta^3  \log \frac{1}{\eta} \gamma_{ij}$.
Then \eqref{newEsingle1} becomes
\begin{eqnarray}
\eta \Bigg(2 \sqrt{m \pi} K_1^{\frac{1}{2}} + \frac{\Gamma_{11} m^2}{4 \pi} K_1^{-1}   \qquad  \qquad  \qquad  \nonumber \\
+  2 \sqrt{ (1 - m) \pi} K_2^{\frac{1}{2}} + \frac{\Gamma_{22} (1- m)^2}{4 \pi} K_2^{-1} \Bigg).  \nonumber
\end{eqnarray}
With respect to $K_1$ and $K_2$ the above is minimized at 
\begin{eqnarray}
K_1 = \left(\frac{\Gamma_{11}}{4} \right)^{\frac{2}{3}} \frac{m}{\pi}, \; K_2 = \left(\frac{\Gamma_{22}}{4} \right)^{\frac{2}{3}} \frac{1- m}{\pi}. \nonumber
\end{eqnarray}
Consequently the average radii of red and yellow bubbles, are
\begin{eqnarray} \label{R1R2}
r_i =  4^{\frac{1}{3}}  \left ( \log \frac{1}{\eta} \right )^{- \frac{1}{3}}  \gamma_{ii}^{-\frac{1}{3}},  \;
i=1,2,   
\end{eqnarray}
from which \eqref{formulaR1R2} follows.

\begin{figure}
\includegraphics[width=2.7cm]{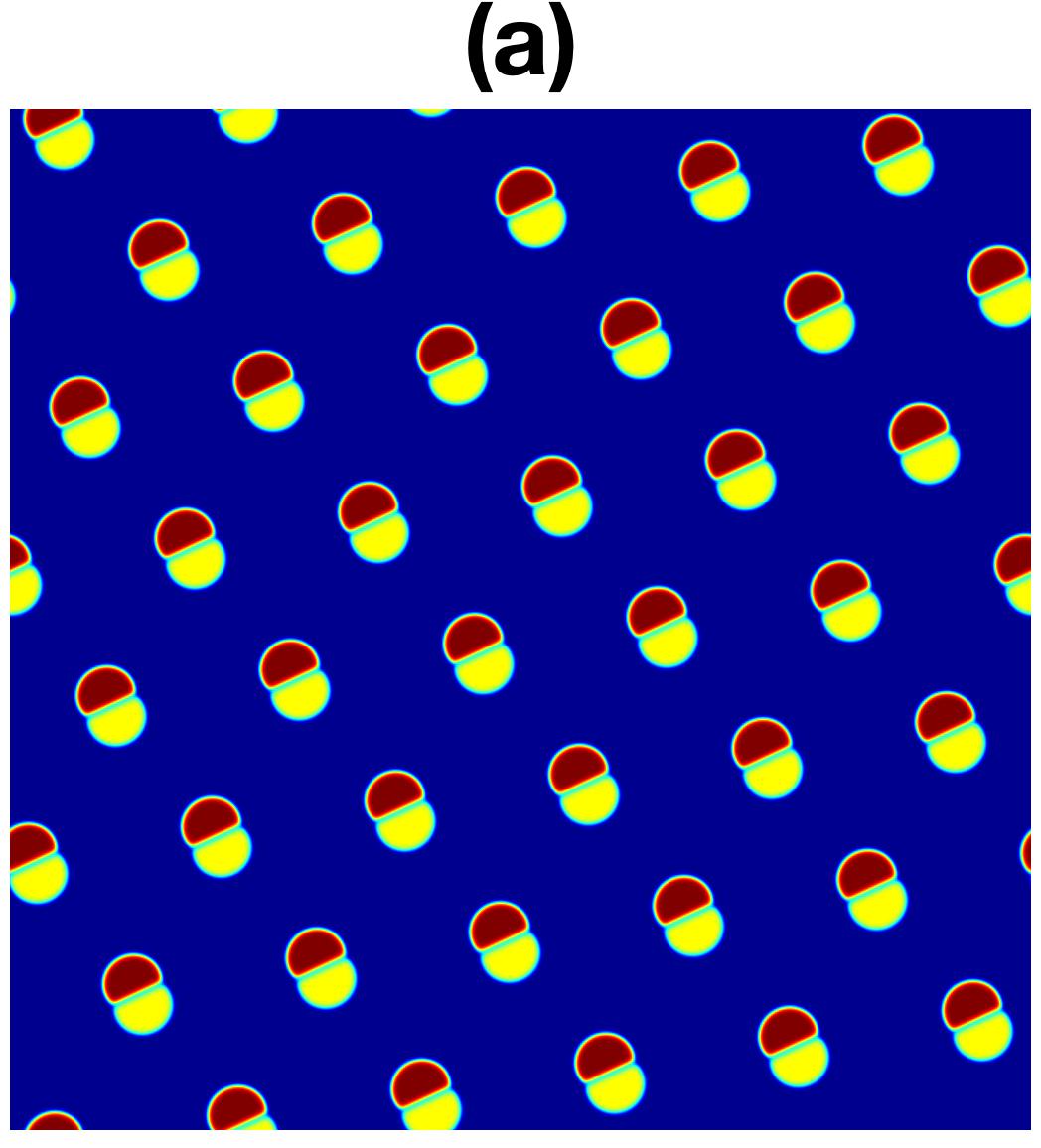} 
\includegraphics[width=2.7cm]{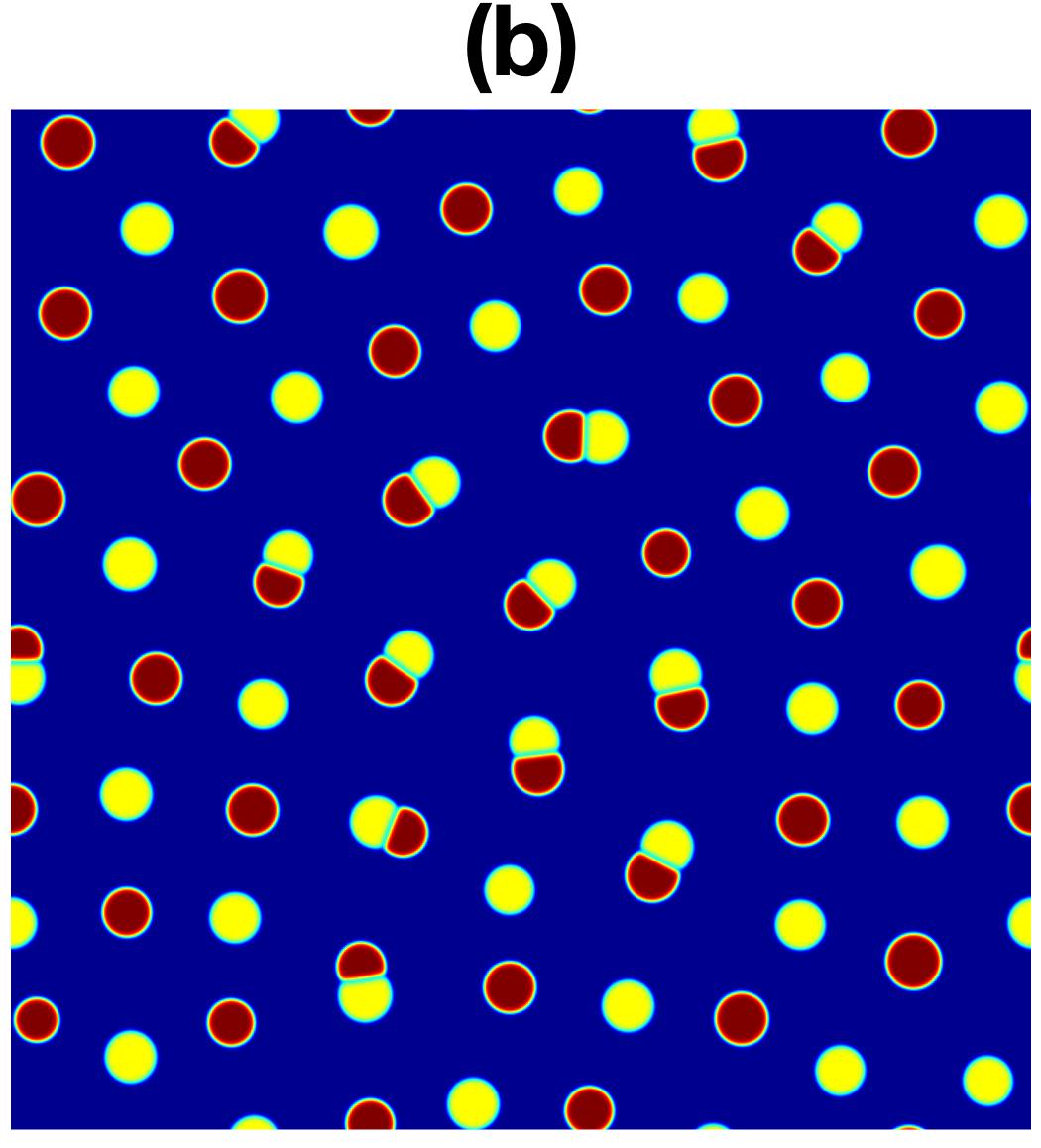} 
\includegraphics[width=2.7cm]{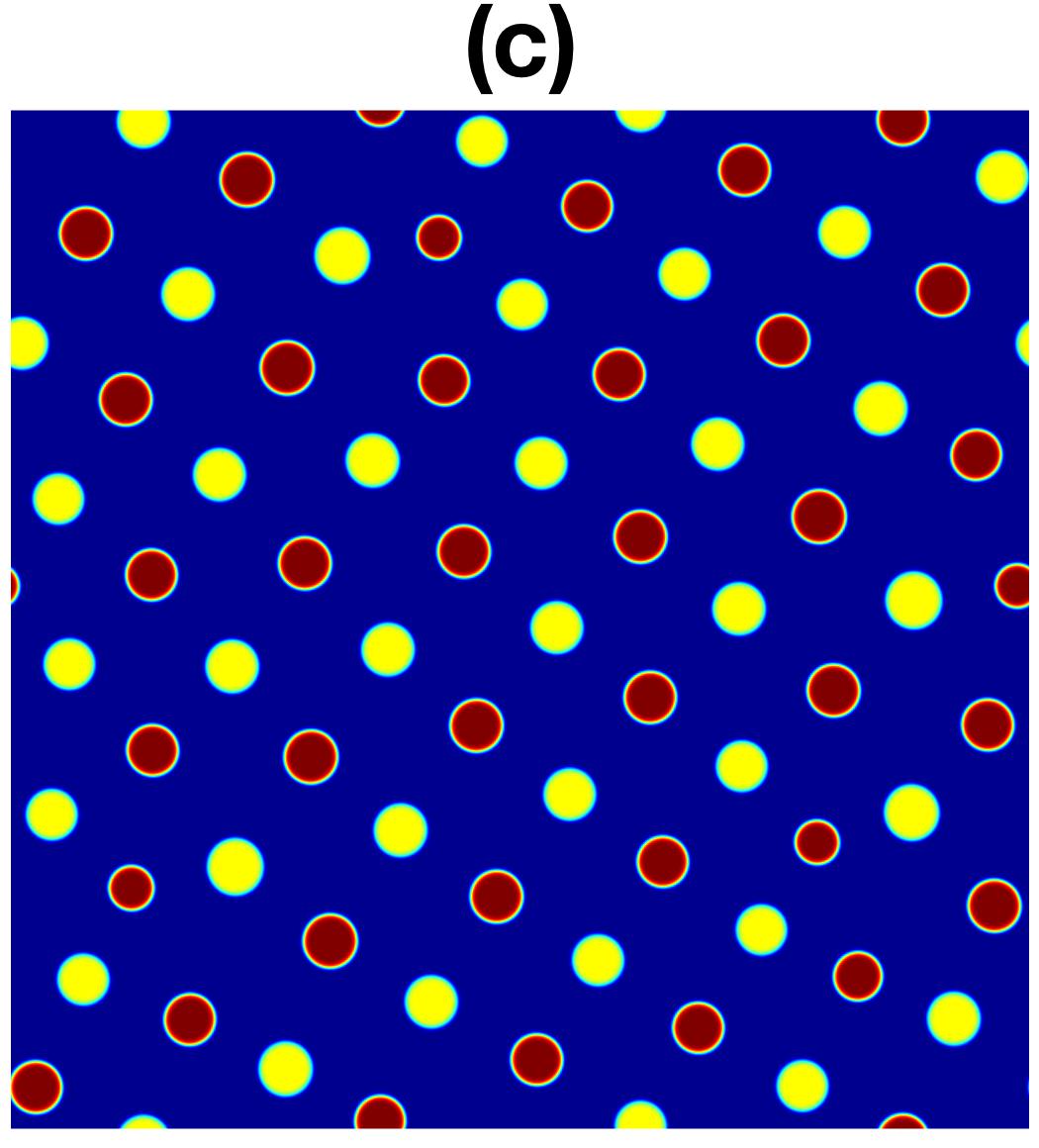} \\
\includegraphics[width=2.7cm]{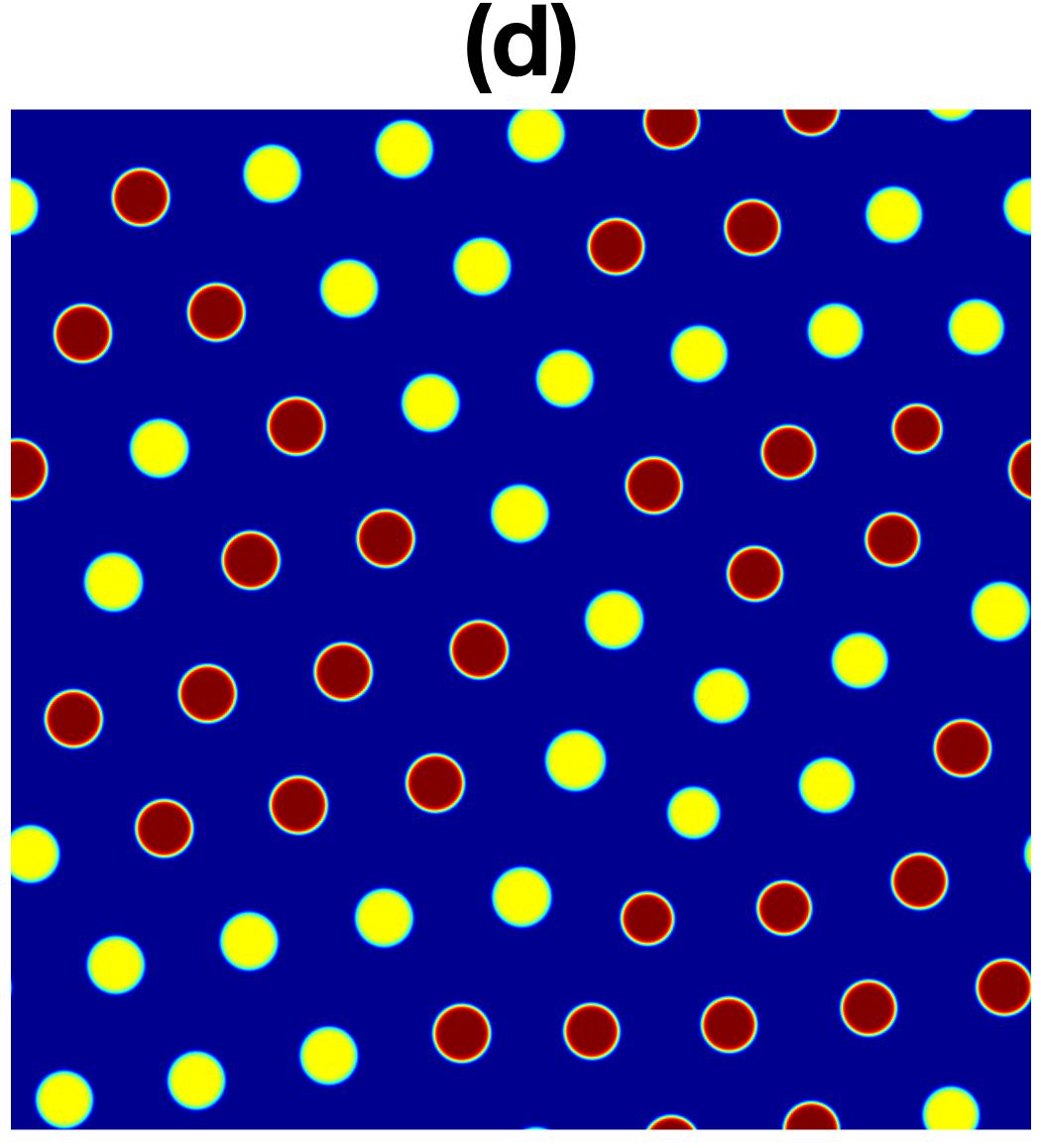} 
\includegraphics[width=2.7cm]{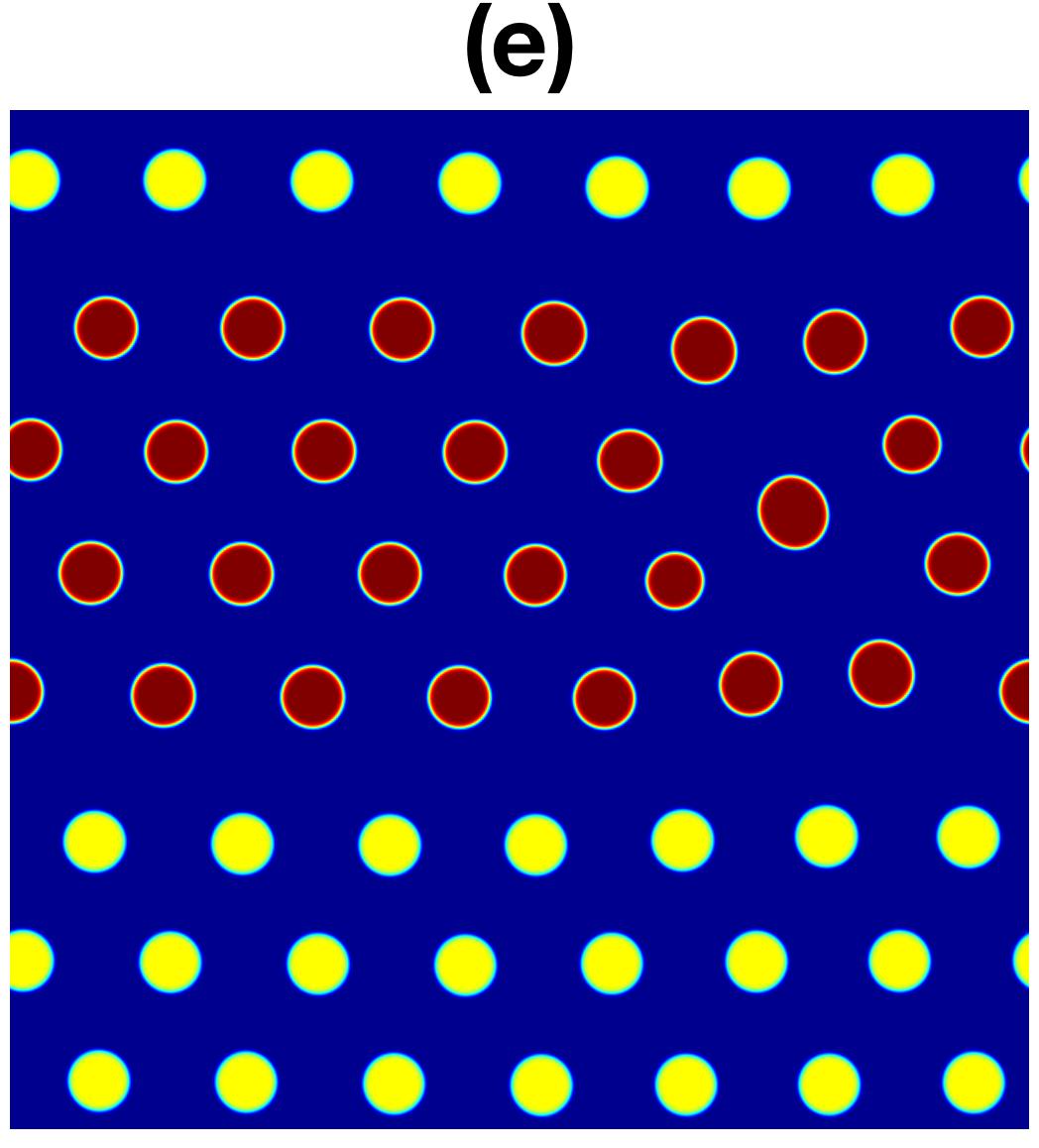} 
\includegraphics[width=2.7cm]{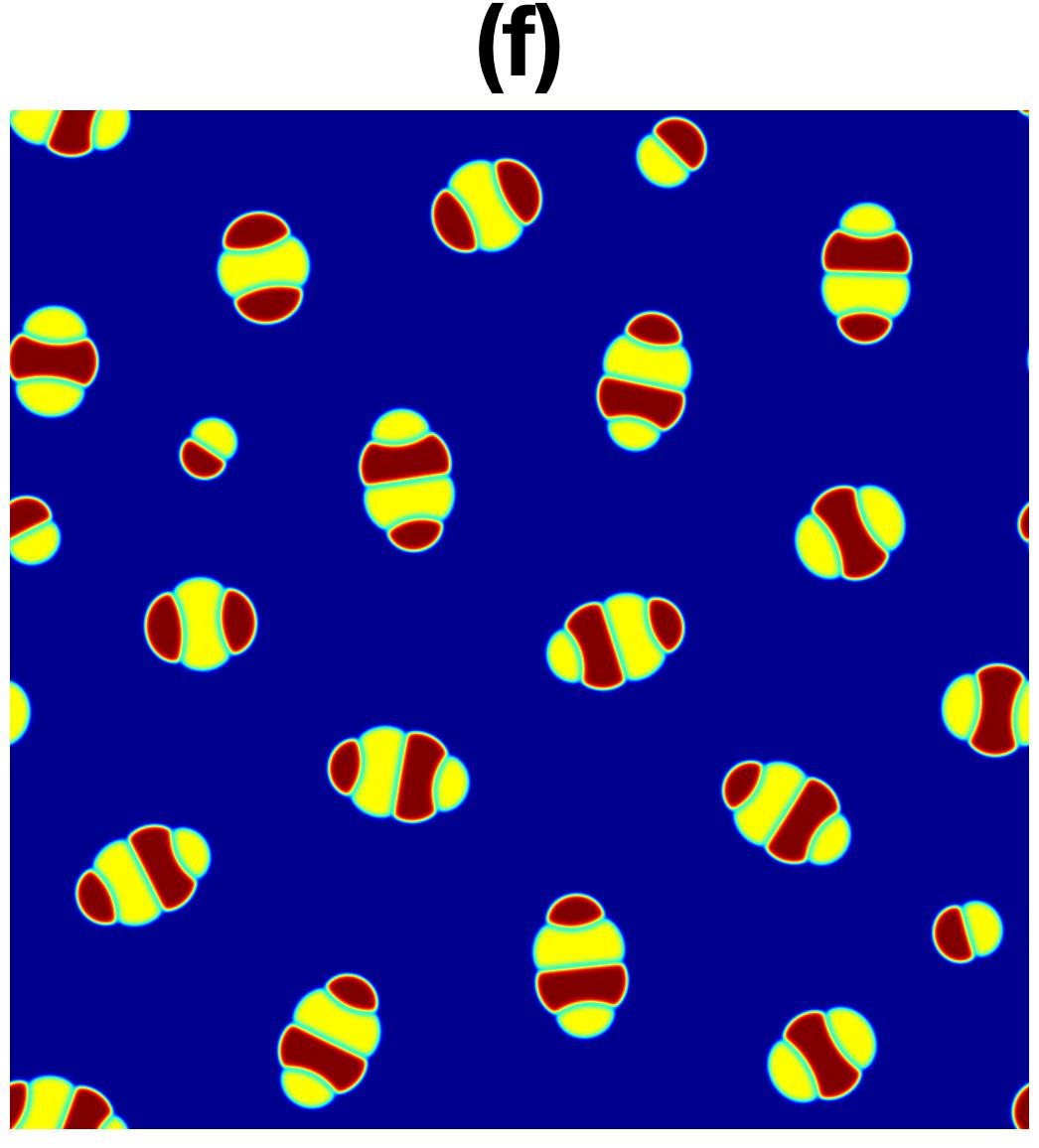} \\
\includegraphics[width=9.1cm]{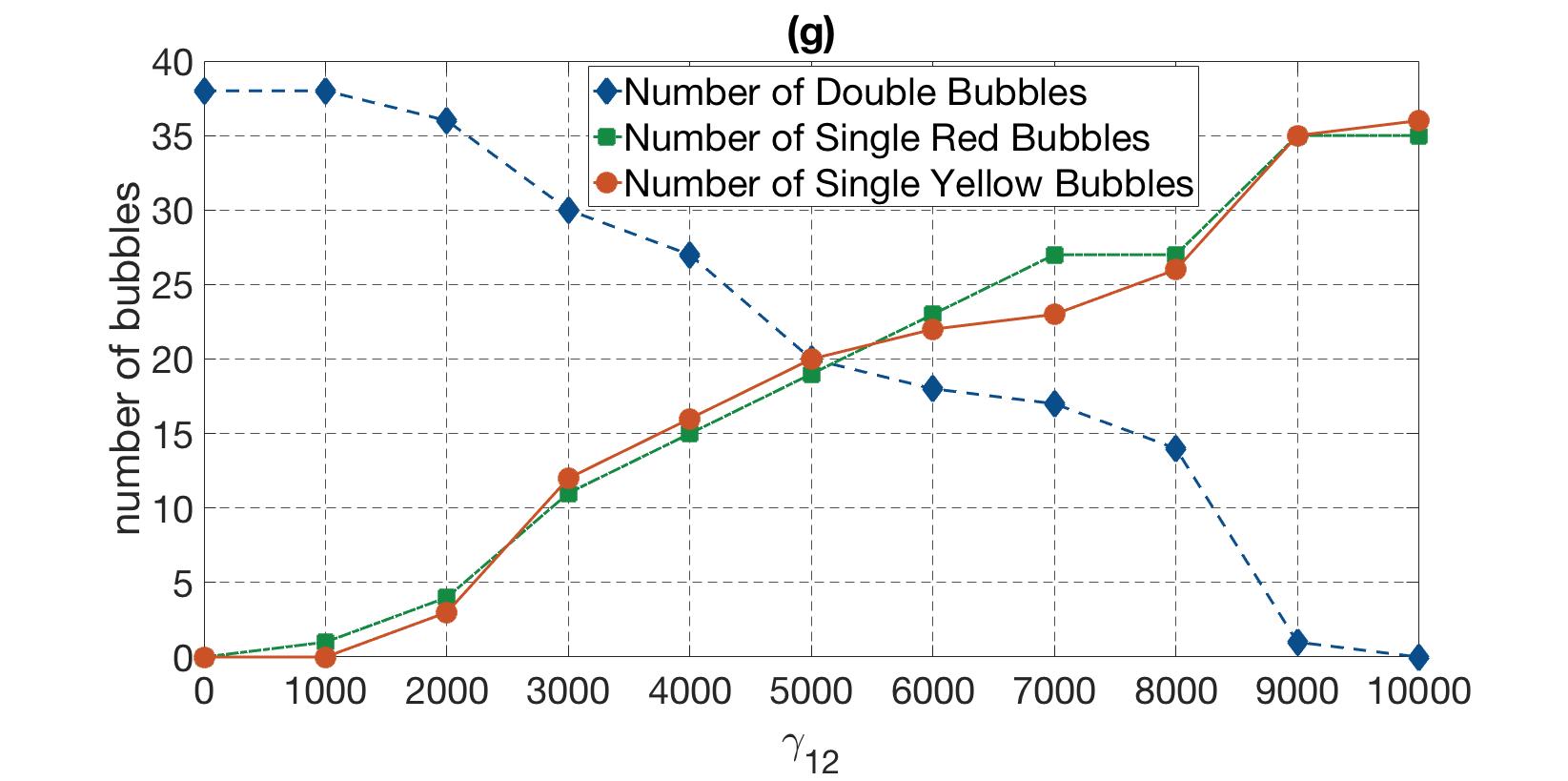} 
\caption{The effect of $\gamma_{12}$. As $\gamma_{12}$ increases, double bubble assemblies change to coexisting single and double bubbles,  and then to single bubble assemblies. 
When $\gamma_{12}$ is negative, nonstandard double bubbles appear.
(a) $\gamma_{12}=0 $, (b) $\gamma_{12}=8,000 $, (c) $\gamma_{12}=10,000 $, (d) $\gamma_{12}=20,000 $, (e) $\gamma_{12}=22,000 $, and (f) $\gamma_{12}= -13,000 $. (g) The numbers of bubbles as $\gamma_{12}$ increases from $0$ to $10,000$.
The other parameters are $\gamma_{11} = \gamma_{22} = 20,000$, $\omega_1 = \omega_2 = 0.09$. }
\label{fig4}
\end{figure}

{\it{Double bubble assemblies.}}---In some parameter ranges, ternary systems may display double bubble assemblies (see Fig. \ref{fig1}  (a)). 
Let $\omega_1 = \omega_2=0.09$, $\gamma_{12}=0$,
and increase $\gamma_{11} = \gamma_{22} $
from $200$ to $40,000$. The number of double bubbles $K_b$ in an assembly increases correspondingly as seen in the insets of Fig. \ref{fig3} (a). The increment of $K_b$ obeys the law $K_b \sim \gamma_{11}^{2/3}$. This confirms that the long range interaction favors small domains.

This two-thirds law can be verified theoretically for both symmetric $(\omega_1 = \omega_2)$ and asymmetric $(\omega_1 \neq \omega_2)$ double bubble assemblies. To this end, consider the strong segregation limit of the free energy $E$ \cite{minimal}.
In an equilibrium state, all double bubbles have approximately the same shape and size. Since each double bubble is bounded by three arcs, let 
$r_i$, $i = 0, 1, 2, $ denote the radii of these arcs and $a_i$, $i=0,1,2$, denote the angles associated with these arcs. Based on \cite{double}, up to the leading order, the free energy is 
\begin{eqnarray} \label{newE2}
  K_b  \left ( \sum_{i=0}^2  a_i r_i +  \sum_{i, j =1}^2  \frac{\gamma_{ij} }{ 4 \pi } \left(  \eta^4 \log \frac{1}{\eta}  \right)  \frac{m_i m_j }{K_b^2} \right),
\end{eqnarray}
where $\eta^2 m = \omega_1 |\Omega|$, $\eta^2 (1-m) = \omega_2 |\Omega|$, $m_1=m$, and $m_2=1-m$.
If $l_i$, $i = 0, 1, 2$, are the radii of the three arcs of a double bubble whose two areas are $m$ and $1-m$,
 then $r_i = \eta l_i/\sqrt{K_b}$. Let $\Gamma_{ij} = \eta^3  \log \frac{1}{\eta} \gamma_{ij}$ and rewrite \eqref{newE2} as 
\begin{eqnarray} \label{newE3}
\eta \left[    \left( \sum_{i=0}^2  a_i l_i \right ) \;  K_b^{\frac{1}{2}} + \left(  \sum_{i, j =1}^2  \frac{\Gamma_{ij} m_i m_j }{ 4 \pi } \right) K_b^{-1} \right ]. \nonumber
\end{eqnarray}
With respect to $K_b$, this is minimized at
\begin{eqnarray}
K_b =  \left( \frac{\sum_{i, j =1}^2  \Gamma_{ij} m_i m_j  }{ 2\pi \sum_{i=0}^2  a_i l_i  } \right)^{\frac{2}{3}}. \nonumber
\end{eqnarray}

Fig. \ref{fig3} (b) shows the relationship between $\gamma_{11}/\gamma_{22}$ and $\omega_1/\omega_2$ when double bubble assemblies occur. The vertical green line for each value of $\omega_1/\omega_2$ indicates the range of $\gamma_{11}/\gamma_{22}$ for which double bubble assemblies exist. Beyond this range, ternary systems display other patterns such as coexisting single and double bubbles.  The range becomes wider when $\omega_1/\omega_2$ approaches 1. Taking $\gamma_{11}/\gamma_{22}$ to be the middle value in each range, and plotting it with respect to the ratio $\omega_1/\omega_2$, one finds that it agrees with the graph of
$y=x^{-3/2}$.

{\it{The effect of $\gamma_{12}$.}}--- As $\gamma_{12}$ increases from $0$, red and yellow constituents tend to break.  In Fig. \ref{fig4}(a), $\gamma_{12} = 0$ and all components are double ones. In Fig. \ref{fig4}(b), $\gamma_{12} = 8,000$,  many double bubbles break into single red and yellow bubbles to yield a coexisting pattern. In Fig. \ref{fig4}(c), $\gamma_{12} = 10,000$, all double bubbles disappear, the assembly becomes a pure single bubble one. In this case the red and yellow bubbles
 are well mixed in an organized way.
 In Fig. \ref{fig4}(d), $\gamma_{12} = 20,000$, the system still displays a single bubble assembly, but the  red and yellow bubbles are mixed randomly; many single bubbles of the same color gather together.
 When $\gamma_{12}=22,000$ is even larger in Fig. \ref{fig4}(e), red bubbles are completely separated from yellow bubbles in the assembly. Note that as $\gamma_{12}$ increases, the matrix $\gamma$ changes from being positive definite, to
 semi-positive definite, and to indefinite.
 In Fig. \ref{fig4} (f), a negative $\gamma_{12}$ is used. Red and yellow constituents tend to be more ``adhesive". Nonstandard double bubbles appear in the assembly.
 In Fig. \ref{fig4} (g), the numbers of single and double bubbles when $\gamma_{12}$ changes from $0$ to $10,000$ are recorded. 
The existence of double bubble assemblies and single bubble assemblies have
been theoretically established recently \cite{double, discAssemblies}.
There have been no theoretical studies on assemblies of coexisting single
and double bubbles or on assemblies of nonstandard double bubbles.

{\it{Conclusion.}}---A computational model is used to study pattern formation
in ternary systems. 
Numerical simulations answered one open question from the theoretical study of triblock copolymers: the polarity direction of double bubbles in double bubble assemblies. It is shown that the average size of red/yellow bubbles in a single bubble
assembly does not depend on $\omega_1/\omega_2$,
the ratio of the area fractions of the minority constituents, but rather
on $\gamma_{11}$ and $\gamma_{22}$, as well as $\omega_1 + \omega_2$.
A relationship between $\gamma_{11} / \gamma_{22}$ and $\omega_1/ \omega_2$ exists in order to have double bubble assemblies. 

This work can be extended in a number of directions. Morphological patterns
in three dimensions can be studied by the same model.
It can also be generalized for quaternary systems, such as tetrablock
copolymers. Other gradient flows of $E$, such as a $H^{-1}$ flow, are also
worth studying.

X.R. is supported by National Science Foundation, DMS-1714371. Y.Z. is supported by a grant from the Simons Foundation through Grant No. 357963.


\begin{thebibliography}{1}
\bibitem{block} F. S. Bates and G. H. Fredrickson, \href{http://physicstoday.scitation.org/doi/10.1063/1.882522}{Phys. Today, {\bf{52}}, No.2, 32 (1999)}.
\bibitem{developments} I.W. Hamley, \textit{Developments in Block Copolymer Science and Technology}, \href{}{(Wiley, New York, 2004)}.
\bibitem{optoelectronics} I. Botiz and S.B. Darling, \href{https://www.sciencedirect.com/science/article/pii/S1369702110700833}{Mater. Today, {\bf{13}}, 42 (2010)}.
\bibitem{takenaka} M. Takenaka {\it{et al.}}, \href{http://pubs.acs.org/doi/abs/10.1021/ma070739u}{Macromolecules, {\bf{40}}, 4399 (2007)}.
\bibitem{blockCoTheo4} E. Helfand and Z.R. Wasserman, \href{http://pubs.acs.org/doi/abs/10.1021/ma60054a001}{Macromolecules, {\bf{9}}, 879 (1976)}.
\bibitem{stable} M.W. Matsen and M. Schick, \href{https://journals.aps.org/prl/abstract/10.1103/PhysRevLett.72.2660}{Phys. Rev. Lett., {\bf{72}}, 2660 (1994)}.
\bibitem{drolet} F. Drolet and G. H. Fredrickson, \href{https://journals.aps.org/prl/abstract/10.1103/PhysRevLett.83.4317}{Phys. Rev. Lett., {\bf{83}}, 4317 (1999)}.
\bibitem{catyler} C. A. Tyler, J. Qin, F. S. Bates and D. C. Morse, \href{http://pubs.acs.org/doi/abs/10.1021/ma062778w}{Macromolecules, {\bf{40}}, 4654 (2007)}.
\bibitem{discoveringOrd} Z. Guo {\it{et al.}}, \href{https://journals.aps.org/prl/abstract/10.1103/PhysRevLett.101.028301}{Phys. Rev. Lett., {\bf{101}}, 028301 (2008)}.
\bibitem{nucleationOrd} X. Cheng {\it{et al.}}, \href{https://journals.aps.org/prl/abstract/10.1103/PhysRevLett.104.148301}{Phys. Rev. Lett., {\bf{104}}, 148301 (2010)}.
\bibitem{jiang} Y. Jiang and J. Z. Y. Chen, \href{https://journals.aps.org/prl/abstract/10.1103/PhysRevLett.110.138305}{Phys. Rev. Lett., {\bf{110}}, 138305 (2013)}.
\bibitem{Oono} Y. Oono and Y. Shiwa, \href{http://www.worldscientific.com/doi/abs/10.1142/S0217984987000077}{Mod. Phys. Lett. B, {\bf{1}}, 49 (1987)}.
\bibitem{cell} M. Bahiana and Y.Oono, \href{https://journals.aps.org/pra/abstract/10.1103/PhysRevA.41.6763}{Phys. Rev. A, {\bf{41}}, 6763 (1990)}.
\bibitem{cellDy} S.R. Ren and I.W. Hamley, \href{http://pubs.acs.org/doi/abs/10.1021/ma000678z}{Macromolecules, {\bf{34}}, 116 (2001)}.
\bibitem{phase-field} X.-F. Wu and Y.A. Dzenis, \href{https://pdfs.semanticscholar.org/9390/ad2aef04b620e785a49f1b25a3ae41b8e4b8.pdf}{Phys. Rev. E, {\bf{77}}, 031807 (2008)}.
\bibitem{equilibrium} T. Ohta and K. Kawasaki, \href{http://pubs.acs.org/doi/abs/10.1021/ma00164a028}{Macromolecules, {\bf{19}}, 2621 (1986)}.
\bibitem{microphase} H. Nakazawa and T. Ohta, \href{http://pubs.acs.org/doi/abs/10.1021/ma00072a031}{Macromolecules, {\bf{26}}, 5503 (1993)}.
\bibitem{kineticPath} S. Qi and Z.-G. Wang, \href{https://journals.aps.org/prl/abstract/10.1103/PhysRevLett.76.1679}{Phys. Rev. Lett.,  {\bf{76}}, 1679 (1996)}.
\bibitem{some} Y. Nishiura and I. Ohnishi, \href{http://www.sciencedirect.com/science/article/pii/016727899500005O}{Physica D, {\bf{84}}, 31 (1995)}.
\bibitem{ontheMulti} X. Ren and J. Wei, \href{http://epubs.siam.org/doi/abs/10.1137/S0036141098348176}{SIAM J. Math. Anal., {\bf{31}}, 909 (2000)}.
\bibitem{onthe} R. Choksi and X. Ren, \href{https://link.springer.com/article/10.1023/A:1025722804873}{J. Statist. Phys., {\bf{113}}, 151 (2003)}.
\bibitem{variation} R. Choksi and P. Sternberg, \href{https://www.degruyter.com/view/j/crll.2007.2007.issue-611/crelle.2007.074/crelle.2007.074.xml}{J. Reine Angew. Math, {\bf{611}}, 75 (2007)}.
\bibitem{dropletPhases} C. B. Muratov, \href{https://link.springer.com/article/10.1007/s00220-010-1094-8}{Comm. Math. Phys., {\bf{299}}, 45 (2010)}.
\bibitem{reorientation} S. Orizaga and K. Glasner, \href{https://journals.aps.org/pre/abstract/10.1103/PhysRevE.93.052504}{Phys. Rev. E, {\bf{93}}, 052504 (2016)}.
\bibitem{aPreconditioner}  P.E. Farrell and J.W. Pearson, \href{http://epubs.siam.org/doi/abs/10.1137/16M1065483}{SIAM J. Matrix Anal. Appl, {\bf{38}}, 217 (2017)}.
\bibitem{amixed} A. Aristotelous, O. Karakashian and S.M. Wise, \href{https://www.aimsciences.org/journals/displayArticles.jsp?paperID=8984}{Discrete Contin. Dyn. Syst. Ser. B, {\bf{18}}, 2211 (2013)}.
\bibitem{efficient} Q. Cheng, X. Yang and J. Shen, \href{http://www.sciencedirect.com/science/article/pii/S0021999117302814?via\%3Dihub}{J. Comput. Phys., {\bf{341}}, 44 (2017)}.
\bibitem{bohbot} Y. Bohbot-Raviv and Z.-G. Wang, \href{https://journals.aps.org/prl/abstract/10.1103/PhysRevLett.85.3428}{Phys. Rev. Lett., {\bf{85}}, 3428 (2000)}.
\bibitem{morphologyABC} W. Zheng and Z.-G. Wang, \href{http://pubs.acs.org/doi/abs/10.1021/ma00125a026}{Macromolecules, {\bf{28}}, 7215 (1995)}.
\bibitem{coreshell} X. Ren and C. Wang, \href{https://aimsciences.org/journals/displayArticlesnew.jsp?paperID=13394}{Discrete Contin. Dyn. Syst., {\bf {37}}, 983 (2017)}.
\bibitem{triblockCo} X. Ren and J. Wei, \href{http://www.sciencedirect.com/science/article/pii/S0167278902008084}{Physica D, {\bf{178}}, 103 (2003)}.
\bibitem{blend} R. Choksi and X. Ren, \href{http://www.sciencedirect.com/science/article/pii/S0167278905001016}{Physica D, {\bf{203}}, 100 (2005)}.
\bibitem{newPhasef} Y. Zhao, Y. Ma, H. Sun,  B. Li and Q. Du, Submitted, (2017). 
\bibitem{multiplier} M. R. Hestenes, \href{https://link.springer.com/article/10.1007/BF00927673}{J. Opti. Theo. Appl., {\bf{4}}, 302 (1969)}.
\bibitem{efficientSta} X. Wang, L. Ju and Q. Du, \href{http://www.sciencedirect.com/science/article/pii/S0021999116300365}{J. Comput. Phys., {\bf{316}}, 21 (2016)}.
\bibitem{applications} L. Q. Chen and J. Shen, \href{http://www.sciencedirect.com/science/article/pii/S001046559700115X}{Comput. Phys. Commun., {\bf{108}}, 147 (1998)}.
\bibitem{coarsening} J. Zhu, L.Q. Chen, J. Shen and V. Tikare, \href{https://journals.aps.org/pre/abstract/10.1103/PhysRevE.60.3564}{Phys. Rev. E, {\bf{60}}, 3564 (1999)}.
\bibitem{double} X. Ren and J. Wei, \href{https://link.springer.com/article/10.1007/s00205-014-0798-x}{Arch. Rat. Mech. Anal., {\bf{215}}, 967 (2015)}. 
\bibitem{molecularCon} T. Hashimoto, H. Tanaka and H. Hasegawa, \href{https://www.elsevier.com/books/molecular-conformation-and-dynamics-of-macromolecules-in-condensed-systems/nagasawa/978-0-444-42993-3}{M. Nagasawa Ed., Elsevier Science, 1988}.
\bibitem{blockCoTheo6} E. Helfand and Z.R. Wasserman, \href{http://pubs.acs.org/doi/abs/10.1021/ma60076a045}{Macromolecules, {\bf{13}}, 994 (1980)}.
\bibitem{theory} L. Leibler, \href{http://pubs.acs.org/doi/abs/10.1021/ma60078a047}{Macromolecules, {\bf{13}}, 1602 (1980)}.
\bibitem{appliModular} X. Chen and Y. Oshita, \href{https://link.springer.com/article/10.1007/s00205-007-0050-z}{Arch. Rat. Mech. Anal., {\bf{186}}, 109 (2007)}.
\bibitem{many} X. Ren and J. Wei, \href{http://www.worldscientific.com/doi/abs/10.1142/S0129055X07003139}{Rev. Math. Phys., {\bf{19}}, 879 (2007)}.
\bibitem{squarePacking} C. Tang {\it{et al.}}, \href{http://pubs.acs.org/doi/abs/10.1021/ma800207n}{Marcromolecules, {\bf{41}}, 4328 (2008)}.
\bibitem{pingtang} P. Tang, F. Qiu, H. Zhang and Y. Yang, \href{https://journals.aps.org/pre/abstract/10.1103/PhysRevE.69.031803}{Phys. Rev. E, {\bf{69}}, 031803 (2004)}.
\bibitem{minimal} S. Baldo, \href{http://www.sciencedirect.com/science/article/pii/S0294144916303043}{Annales de l'I.H.P., {\bf{7}}, 67 (1990)}.
\bibitem{discAssemblies} X. Ren and C. Wang, Submitted, (2017).
\end{thebibliography}
\end{document}